\documentclass[aps,pra,twocolumn,groupedaddress,showpacs]{revtex4-1}

\usepackage{amsmath}
\usepackage{graphicx}
\usepackage{latexsym,amssymb,epsf}

\begin{document}


\title{Probing the Dynamical Behaviour of Surface Dipoles Through Energy Absorption
Interferometry}

\author{S. Withington}
\email{stafford@mrao.cam.ac.uk}
\author{C.N. Thomas}
\affiliation{Cavendish Laboratory, JJ Thompson Avenue, Cambridge, UK}

\date{\today}

\begin{abstract}
Spatial interferometry, based on the measurement of total absorbed power, can be used to determine the state of coherence of the electromagnetic field to which any energy-absorbing structure is sensitive. The measured coherence tensor can be diagonalized to give the amplitude, phase, polarization patterns, and responsivities of the individual electromagnetic modes through which the structure can absorb energy. Because the electromagnetic modes are intimately related to dynamical modes of the system, information about collective excitations can be found. We present simulations, based on the Discrete Dipole Approximation (DDA), showing how the dynamical modes of systems of surface dipoles can be recovered. Interactions are taken into consideration, leading to long-range coherent phenomena, which are revealed by the method. The use of DDA enables the interferometric response of a wide variety of objects to be modeled, from patterned photonic films to biological macromolecules.
\end{abstract}



\maketitle

\section{INTRODUCTION} \label{sec:introduction}

Understanding the processes by which molecules and particles on surfaces absorb and scatter electromagnetic radiation is an area of considerable scientific importance. Information may be needed because of an intrinsic desire to study surface physics, because an active surface forms part of an electronic device, or because of the need to locate defects and identify contaminants. Examples include the dynamics of adsorbed biomolecules \cite{Long_a}, the adhesion and movement of droplets \cite{Li_a}, the plasmonic behaviour of nanoparticles \cite{Shegai_a, Zheng_a}, and the photonic response of patterned films \cite{Greffet_a, Wadsworth_a}. Surface processes have traditionally been studied through infrared and optical microscopy, spectroscopy, and scattering, with AFM playing a key role in recent years.

Often, the information needed has to be acquired indirectly through measurements made at inappropriate wavelengths, or through models that are hard to extrapolate. Crucially, it is necessary to measure the state of coherence of the field to which a system is sensitive in order to fully characterize its behaviour: for example when calculating radiative heat transfer between nanoparticles \cite{Rousseau_a, Biehs_a}. Measurements made with a  single source only gain access to a subset of the information available. Of particular importance are those cases where the ability of a surface to absorb energy is central to the operation of a device. For example, complex schemes are being explored to increase the efficiency of photovoltaics \cite{Pillai_a, Luque_a}, to optimize the sensitivity of infrared and optical sensors \cite{Thomas_a}, and to develop frequency selective surfaces, high-power mirrors, and hyperspectral black absorbers. In all cases, the physics of the absorption process is intimately related to the existence of coherent phenomena such as resonant dipole interactions, surface plasmon coupling \cite{Setala_a}, and surface phonon exchange. Even in the case of apparently simple structures, such as planar dielectric and metallic interfaces, the near-field spatial and polarimetric correlations contain a wealth of information \cite{Henkel_a, Joulain_a, Lau_a}.

A typical set of questions is as follows: What are the scale sizes and forms of the dynamical modes responsible for absorbing energy? How many modes are involved in the absorption process? What are their responsivities? What physical mechanisms mediate the collective behaviour? In this paper we argue that it is possible to measure the state of coherence of the electromagnetic field to which any system is sensitive, and thereby determine the spatial forms, polarizations, and relative  strengths of the individual modes responsible for absorbing energy. Because the electromagnetic modes are intimately related to dynamical modes of the system, it is possible to accrue direct information about the dynamical modes themselves, and thereby reveal the existence of collective phenomena. The basic technique can be used at any wavelength, and can be implemented in a number of different ways.

The basic idea is to illuminate the Structure Under Test (SUT) with a pair of phase-locked, near-field or far-field point sources. As the differential phase between the sources is rotated, the {\em total} absorbed power displays a fringe. If the complex visibility of this fringe is measured for different pairs of source locations, the dynamical forms of the modes responsible for absorbing the energy can be found. In the case of detectors and photovoltaics, the scheme is particularly easy to implement because the degrees of freedom responsible for generating the measured output are those of practical significance.

Although the technique is applicable to a wide range of problems, we illustrate the method by showing how it provides information about the collective behaviour of systems of surface dipoles. The SUT is modeled by using the Discrete Dipole Approximation (DDA) \cite{Chaumet_a, Draine_a, Yurkin_a}, together with a polarizability  that allows for bound states and damping  to the substrate. DDA was originally devised \cite{Draine_b} to model light extinction by astrophysical particles, but has since been developed in areas as diverse as modeling the aggregation of irregular particles from colloids \cite{Felidj_a}, the scattering of light by particles on surfaces \cite{Taubenblatt_a, Litz_a}, the scattering of light by periodic targets \cite{Draine_c}, and the optical properties of viruses, proteins, DNA, and blood cells \cite{Shermila_b, Umazano_a, Yurkin_b}. Formulating our measurement procedure in terms of DDA opens the way to predicting the interferometric response of a wide variety of complex objects such as biomolecules, droplets, and nanoparticles, and thereby creating representative dynamical models of measured interferometric data. Other simulation methods, such as MoM, can be used.

As an aside, a variant of EAI can be used with DDA purely as a numerical tool for determining the state of coherence, and therefore the incident electromagnetic modes, to which any irregular dielectric body is sensitive.
The scheme differs from the usual method for determining the state of coherence of thermally radiated fields, through
\begin{equation}
\label{A}
\overline{\overline{W}}({\bf r},{\bf r}') = \mu_{0} \hbar \omega^{2} \coth \left( \frac{\hbar \omega}{2 k_{B} T} \right) \mbox{Im} \, \overline{\overline{G}}^{e}({\bf r},{\bf r}')
\mbox{,}
\end{equation}
where $\overline{\overline{G}}^{e}({\bf r},{\bf r}')$ is the electric Green's dyadic of the whole system including scattering, because (\ref{A}) requires the object to be in thermal equilibrium with its surroundings, and therefore includes the blackbody environment and scattered fields. As a consequence, even lossless dielectric objects display near-field coherence effects, as discussed by Yannopapas in the context of chains of spheres \cite{Yannopapas_a}. In contrast, EAI only includes those degrees of freedom responsible for absorbing energy.

In section~\ref{sec:basic_scheme} we present the theory of Energy Absorption Interferometry (EAI), and then in
section~\ref{sec:simulations} illustrate how it can be used to recover the dynamical modes of ring-like and linear chains of coupled dipoles, with and without defects. Linear chains of dipoles \cite{Markel_a}, spheres \cite{Yannopapas_a}, and cylinders \cite{Lindberg_a} are of practical importance, and exhibit complex forms of spatial and spectral coherence. We show how EAI can reveal large-scale collective behaviour, and how the response of a discrete system transitions to that of a continuous system when the SUT contains many dipoles per wavelength. In section~\ref{sec:practical}, we discuss practical implementation.

\begin{figure}[!]
\begin{center}
\includegraphics[scale=0.7]{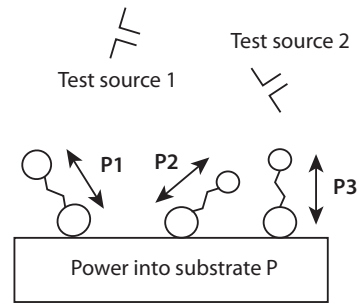}
\caption{\label{fig1} A collection of particles on a surface. The individual particles have dipole moments ${\bf p}_{i}$. The system is illuminated by a pair of phase-locked sources, which can have any polarisation and position. The experiment measures the complex visibility of the absorbed power as the differential phase between the sources is varied.}
\end{center}
\end{figure}

\section{Basic Scheme}\label{sec:basic_scheme}

\subsection{Dynamical Behaviour}

Consider a collection of particles whose electromagnetic behaviour can be modeled as a system of dipoles. Although EAI can be applied to any three-dimensional distribution, assume that the particles are bound to a surface, Fig.~\ref{fig1}. The dynamical response of dipole $i$ can be described by the Lorentz model
\begin{equation}
\label{A1}
\frac{\partial^{2} {\bf x}_{i} (t)}{\partial t^{2}} + \Gamma \frac{\partial {\bf x}_{i} (t)}{\partial t} + \omega_{oi}^{2} {\bf x}_{i} (t) = \frac{q_{i}}{m_{i}} {\bf E}({\bf r}_{i},t) \cdot \hat{\bf x}_{i}  \hat{\bf x}_{i}
\mbox{,}
\end{equation}
where ${\bf x}_{i}$ is the effective separation vector of charge $q_{i}$ from its partner pinned at ${\bf r}_{i}$. The dyadic product $\hat{\bf x}_{i}  \hat{\bf x}_{i}$ constrains $q_{i}$ to move in the $\hat{\bf x}_{i}$ direction, but other dyads could be used to place more general constraints on the degrees of freedom of motion. $\omega_{oi}$ is the resonant frequency due to binding, $m_{i}$ the mass, and $\Gamma_{i}$ the momentum relaxation rate of dipole $i$. The damping arises from a transfer of energy to the phonon or electron system of the substrate. (\ref{A1}) does not include a radiation-reaction term, because we are only interested in the absorbed power, and the radiation-reaction term has little effect on behaviour. ${\bf E}({\bf r}_{i},t)$ is the applied field, which includes the incident field, ${\bf E}^{I}({\bf r}_{i},t)$, and the scattered field from all other dipoles, ${\bf E}^{S}({\bf r}_{i},t)$. In the parlance of DDA, ${\bf E}({\bf r}_{i},t)$ is the `local field' because it does not include the self field of the dipole at ${\bf r}_{i}$. Invoking time-harmonic behaviour, $e^{-i \omega t}$, leads to a polarization of
\begin{equation}
\label{A2}
{\bf P}( {\bf r} ) = \sum_{i} \frac{q_{i}^{2}}{m_{i}} \frac{\delta ( {\bf r} - {\bf r}_{i} )}{\left( \omega_{oi}^{2} - \omega^{2} - i \omega \Gamma_{i} \right)}{\bf E}({\bf r}_{i}) \cdot \hat{\bf x}_{i}  \hat{\bf x}_{i}
\mbox{.}
\end{equation}
The anisotropic susceptibility tensor, $\overline{\overline{\mathbf \chi}}_{i} ( {\bf r} )$, defined through ${\bf P}( {\bf r} ) = \epsilon_{o} \overline{\overline{\mathbf \chi}}_{i} ( {\bf r} ) \cdot {\bf E}({\bf r})$, becomes
\begin{equation}
\label{A3}
\overline{\overline{\mathbf \chi}}_{i}( {\bf r} ) = \sum_{i} \frac{q_{i}^{2}}{\epsilon_{o} m_{i}} \frac{\delta ( {\bf r} - {\bf r}_{i} )}{\left( \omega_{oi}^{2} - \omega^{2} - i \omega \Gamma_{i} \right)} \hat{\bf x}_{i}  \hat{\bf x}_{i}
\mbox{.}
\end{equation}
It will be convenient to describe behaviour in terms of polarisation current density ${\bf J}({\bf r}) = \partial {\bf P}(t) / \partial t$, giving ${\bf J}({\bf r}) = \overline{\overline{\sigma}}( {\bf r} ) \cdot {\bf E}({\bf r})$, where
\begin{equation}
\label{A4}
\overline{\overline{\sigma}}( {\bf r} ) = \sum_{i} \frac{q^{2}_{i}}{m_{i}} \frac{-i \omega \, \delta ( {\bf r} - {\bf r}_{i} )}{\left( \omega_{oi}^{2} - \omega^{2} - i \omega \Gamma_{i} \right) } \hat{\bf x}_{i}  \hat{\bf x}_{i}
\mbox{.}
\end{equation}
Large currents can flow at the resonance frequencies, $\omega_{oi}$, and an increase in either binding, $\omega_{oi}$, or damping, $\Gamma_{i}$, decreases the low-frequency conductivity. The permittivity and conductivity tensors are valid in both the high-frequency and quasi-static regimes.

To simulate EAI, it is necessary to calculate the total time-averaged power, $W$, transferred to the substrate. The power can be calculated by treating the system as a dielectric (\ref{A3}), or conductor (\ref{A4}). Adopting the latter,
\begin{eqnarray}
\label{A6}
W  & = &  \frac{1}{2} \mbox{Re} \int {\bf E}^{\ast}({\bf r}) \cdot {\bf J}({\bf r})  \, d^{3}{\bf r} \\ \nonumber
   & = &  \frac{1}{2} \mbox{Re} \int {\bf E}^{\ast}({\bf r})\cdot \overline{\overline{\sigma}}( {\bf r} ) \cdot
   {\bf E}({\bf r})  \, d^{3}{\bf r},
\end{eqnarray}
giving
\begin{equation}
\label{A7}
W = \sum_{i} \frac{q_{i}^{2}}{2 m_{i}} \frac{\Gamma_{i}}{\omega^{2} \left[ \left( \omega_{oi} / \omega \right)^{2} -1
\right]^{2} + \Gamma_{i}^{2}}  \left| {\bf E}({\bf r}_{i}) \cdot \hat{\bf x}_{i} \right|^{2}
\mbox{.}
\end{equation}
It is convenient to present (\ref{A6}) in the symmetric form
\begin{equation}
\label{A8}
 W =  \frac{1}{2} \mbox{Re} \int \int {\bf E}^{\ast}({\bf r}_{1})\cdot \overline{\overline{\sigma}}( {\bf r}_{1},{\bf r}_{2} ) \cdot {\bf E}({\bf r}_{2})  \, d^{3}{\bf r}_{1} d^{3}{\bf r}_{2}
\mbox{,}
\end{equation}
where
\begin{equation}
\label{A9}
\overline{\overline{\sigma}}( {\bf r}_{1},{\bf r}_{2}) = \overline{\overline{\sigma}}( {\bf r}_{1})
\delta( {\bf r}_{1} - {\bf r}_{2} )
\mbox{,}
\end{equation}
is diagonal in position, which emphasizes its local form. When applied to the system of dipoles,
\begin{equation}
\label{A10}
 W=  \frac{1}{2}  \mbox{Re} \sum_{i} \sum_{j} {\bf E}^{\ast}({\bf r}_{i})\cdot \overline{\overline{\sigma}}_{ij} \cdot {\bf E}({\bf r}_{j})
\mbox{,}
\end{equation}
where
\begin{equation}
\label{A11}
\overline{\overline{\sigma}}_{ij} = \frac{q^{2}_{i}}{m_{i}} \frac{-i \omega \, \delta_{ij} }{\left( \omega_{oi}^{2} - \omega^{2} - i \omega \Gamma_{i} \right)} \hat{\bf x}_{i}  \hat{\bf x}_{i}
\end{equation}
is a discretized conductivity tensor.

(\ref{A6}) and (\ref{A8}) cannot be used unless the total electric field, ${\bf E}({\bf r})$ is known, but the total electric field is the sum of the illuminating field, ${\bf E}^{I}({\bf r})$, and the scattered field, ${\bf E}^{S}({\bf r})$, giving
\begin{equation}
\label{A12}
{\bf J}({\bf r}) = \overline{\overline{\sigma}}( {\bf r} ) \cdot \left[ {\bf E}^{I}({\bf r}) + {\bf E}^{S}({\bf r}) \right]
\mbox{.}
\end{equation}
The scattered field can be calculated through the Green's dyadic of the environment,
\begin{equation}
\label{A13}
{\bf E}^{S}({\bf r}) = \int \overline{\overline{\bf G}}( {\bf r}; {\bf r}' ) \cdot {\bf J}({\bf r}') \, d^{3}{\bf r}'
\mbox{,}
\end{equation}
where all multiplicative factors have been included in $\overline{\overline{\bf G}}( {\bf r}; {\bf r}' )$. Green's dyadics of dipoles on, and under, dielectric surfaces have been studied extensively, including multilayers \cite{Sipe_a} and surface roughness \cite{Sano_a}. Ordinarily, we might use the Green's dyadic corresponding to the dielectric half space shown in Fig.~\ref{fig1}, but for illustrative purposes, and consistency with DDA, we shall use the Green's dyadic of free space, ignoring the complex refractive index of the substrate. Of course, the surface could also be modelled through DDA. In the case of particles on very thin, say SiN, membranes, this assumption is valid without further work, but in any case, it does not alter the substance of what follows. The same functional forms shall be used to describe scattering by the dipoles and the excitation of the system by external point sources. The free-space Green's dyadic is
\begin{equation}
\label{A14}
\overline{\overline{\bf G}}({\bf r}; {\bf r}') = \left[ \overline{\overline{\bf I}} + \frac{1}{k_{o}^{2}}
\nabla \nabla \right] g({\bf r}| {\bf r}')
\mbox{,}
\end{equation}
where $g({\bf r} ; {\bf r}')$ is the scalar Green's function
\begin{equation}
\label{A15}
g({\bf r} ; {\bf r}') = \frac{1}{4 \pi |{\bf r} - {\bf r}'|}
\exp \left[ i k_{o} |{\bf r} - {\bf r}'| \right]
\mbox{.}
\end{equation}
$\nabla$ denotes the gradient with respect to the observation point ${\bf r}$. The first term in (\ref{A14}) emerges from the magnetic vector potential, and the second term from the electric scalar potential, and therefore $\overline{\overline{\bf G}}({\bf r} ; {\bf r}')$ includes coupling due to current flow and the exposure of charge.

In order to perform numerical simulations, it is convenient to express $\overline{\overline{\bf G}}({\bf r} ; {\bf r}')$ in Cartesian coordinates. The Green's dyadic has near-field (NF) $R < \lambda$, intermediate-field (IF) $R \sim \lambda$, and far-field (FF)  $R >\lambda$, contributions:
\begin{equation}
\label{A16}
\overline{\overline{\bf G}} = \overline{\overline{\bf G}}_{NF} + \overline{\overline{\bf G}}_{IF} + \overline{\overline{\bf G}}_{FF}
\mbox{.}
\end{equation}
Setting $ {\bf R} = {\bf r}_{j} - {\bf r}_{i}$, and evaluating (\ref{A14}) gives
\begin{eqnarray}
\label{A17}
\overline{\overline{\bf G}}_{NF} & = & - i \omega \mu_{0} \frac{ \exp\left[ i k R \right] }{4 \pi R} \frac{1}{k^{2}R^{2}} \left[ \overline{\overline{\bf I}} - \frac{3 {\bf R}{\bf R}}{R^{2}} \right] \\ \nonumber
\overline{\overline{\bf G}}_{IF} & = & i \omega \mu_{0} \frac{ \exp\left[ i k R \right] }{4 \pi R} \frac{i}{k R} \left[
\overline{\overline{\bf I}} - \frac{3 {\bf R}{\bf R}}{R^{2}} \right] \\ \nonumber
\overline{\overline{\bf G}}_{FF} & = & i \omega \mu_{0} \frac{ \exp\left[ i k R \right] }{4 \pi R} \left[
\overline{\overline{\bf I}} - \frac{{\bf R}{\bf R}}{R^{2}} \right]
\mbox{,}
\end{eqnarray}
where $ R = |{\bf R}|$. By expressing the Green's dyadic in this way, individual terms can be turned on and off during simulations to gain a better understanding of behaviour. It can be shown that as $\omega \rightarrow 0$, $\overline{\overline{\bf G}}_{NF}$ leads to the quasi-static polarization of the dipoles, and therefore electrostatic coupling is included. We do not include the depolarization term $- i ( \omega \mu_{0} /3 k^{2}) \overline{\overline{\bf I}} \delta({\bf R})$, consistent with the use of the localized rather than total field. In DDA, where each dipole represents a small cell of a large dielectric body, care in needed to distinguish between the total and localized fields \cite{Chaumet_a}, particularly if power is dissipated in the dielectric itself. EAI is equally applicable to this case, but for presentational purposes, we have assumed that each particle is small enough that it can be represented by a discrete dipole.

There are several possible ways forward: substitute (\ref{A13}) in (\ref{A12}) and attempt to solve for ${\bf J}({\bf r})$, or substitute (\ref{A12}) in (\ref{A13}) and attempt to solve for ${\bf E}({\bf r})$. The second approach is well conditioned, and so
\begin{eqnarray}
\label{A18}
{\bf E}({\bf r}) & =  & {\bf E}^{I}({\bf r}) + {\bf E}^{S}({\bf r}) \\ \nonumber
 & = & {\bf E}^{I}({\bf r}) + \int \overline{\overline{\bf G}}( {\bf r}; {\bf r}' ) \cdot \overline{\overline{\sigma}}( {\bf r}' ) \cdot {\bf E}({\bf r}') \, d^{3}{\bf r}'
\mbox{.}
\end{eqnarray}
Then
\begin{equation}
\label{A19}
{\bf E}({\bf r}') = \int \left[ \overline{\overline{\bf M}}( {\bf r}; {\bf r}' ) \right]^{-1} \cdot
{\bf E}^{I}({\bf r}) \, d^{3}{\bf r}
\mbox{,}
\end{equation}
where, symbolically, $\left[ \overline{\overline{\bf M}}( {\bf r}; {\bf r}' ) \right]^{-1}$ is the inverse kernel corresponding to
\begin{equation}
\label{A20}
\overline{\overline{\bf M}}( {\bf r}; {\bf r}' ) = \left[  \overline{\overline{\bf I}} \delta ( {\bf r} - {\bf r}' )
- \overline{\overline{\bf G}}( {\bf r}; {\bf r}' ) \cdot \overline{\overline{\sigma}}( {\bf r}' ) \right]
\mbox{.}
\end{equation}

Because only a finite number of dipoles is involved, the inverse operator (\ref{A19}) exists. Substituting
$\overline{\overline{\sigma}}( {\bf r}' )$ in (\ref{A18}), and noting that it is only necessary to calculate the total field at discrete points in order to calculate the dissipated power,
\begin{equation}
\label{A21}
 \sum_{i} \overline{\overline{\bf M}}( {\bf r}_{j}; {\bf r}_{i} ) \cdot {\bf E}({\bf r}_{i}) = {\bf E}^{I}({\bf r}_{j})
\mbox{,}
\end{equation}
where
\begin{equation}
\label{A22}
\overline{\overline{\bf M}}( {\bf r}_{j}; {\bf r}_{i} ) = \left[  \overline{\overline{\bf I}} \delta_{ij}
- \frac{q^{2}_{i}}{m_{i}} \frac{-i \omega}{\left( \omega_{oi}^{2} - \omega^{2} - i \omega \Gamma_{i} \right)}
 \overline{\overline{\bf G}}( {\bf r}_{j}; {\bf r}_{i} ) \cdot \hat{\bf x}_{i}  \hat{\bf x}_{i}
\right]
\mbox{.}
\end{equation}
In DDA, parameters in the Green's and conductivity tensors are combined to express the scattered field in term of a modified Green's tensor and the susceptibility tensor. The physics is, however, the same. It is convenient to write (\ref{A22}) in the form
\begin{equation}
\label{A22b}
\overline{\overline{\bf M}}( {\bf r}_{j}; {\bf r}_{i} ) = \left[  \overline{\overline{\bf I}} \delta_{ij}
- \frac{-i (\alpha_{ei} / \epsilon_{o}) }{\left( 1 - \tilde{\omega_{i}}^{2} - i \tilde{\omega_{i}} \tilde{\Gamma_{i}} \right)} \overline{\overline{\bf G}}'( {\bf r}_{j}; {\bf r}_{i} ) \cdot \hat{\bf x}_{i}  \hat{\bf x}_{i}
\right]
\mbox{,}
\end{equation}
where the tilde denotes normalization with respect to the resonant frequency of the bound dipole: $\tilde{\omega_{i}} = \omega /\omega_{oi}$ and $\tilde{\Gamma_{i}} = \Gamma_{i} /\omega_{oi}$. $\overline{\overline{\bf G}}'( {\bf r}_{j}; {\bf r}_{i} )$ is the Green's dyadic introduced previously, but with the prefactor $\omega \mu_{0}$ changed to $k^{2}$.
According to (\ref{A3}), $\alpha_{ei} = q^{2}_{i} / ( m_{i} \omega_{oi}^{2})$ is the electrostatic, $\omega \rightarrow 0$, polarizability of dipole $i$. The Clausius-Mossotti relationship can be used for simple shapes:
\begin{equation}
\label{A22c}
\frac{\alpha_{ei}}{\epsilon_{o}} = \left[ \frac{(\epsilon_{r}-1)}{1+n(\epsilon_{r}-1)} \right] V
\mbox{,}
\end{equation}
where $V$ is the volume of the particle, having shape factor $0<n<1$.

Because the scheme is fully discretized, it is convenient to introduce matrix notation. If $\mathsf{e}^{I}$ and $\mathsf{e}$ are column vectors containing the complex amplitudes of the incident and total Cartesian field components at the positions of the dipoles, then (\ref{A21}) can be solved: $\mathsf{e} = \mathsf{M}^{-1} \mathsf{e}^{I}$. In matrix notation, (\ref{A10}) becomes
\begin{eqnarray}
\label{A23}
 W  & =  & \frac{1}{2} \, \mbox{Re} \left[ \, \mathsf{e}^{\dagger} \mathsf{\Sigma} \, \mathsf{e} \, \right] \\ \nonumber
  & = & \frac{1}{2} \, \mathsf{e}^{\dagger} \mathsf{\Sigma}^{r}  \mathsf{e} \\ \nonumber
  & = & \frac{1}{2} \, \mathsf{e}^{I \dagger } \left[ \mathsf{M}^{-1} \right]^{\dagger} \mathsf{\Sigma}^{r} \left[ \mathsf{M}^{-1} \right] \mathsf{e}^{I} \\ \nonumber
  & = &  \mathsf{e}^{I \dagger } \mathsf{L} \mathsf{e}^{I}
\mbox{,}
\end{eqnarray}
where $\mathsf{\Sigma}$ is the matrix equivalent of (\ref{A11}), and $\mathsf{\Sigma}^{r}$ is the real part. Taking the trace of each side of the last line of (\ref{A23}), noting that the trace is invariant to cyclic permutation, and calculating the expectation value gives
\begin{equation}
\label{A24}
 \langle W \rangle =  \mbox{Tr} \left[ \langle \mathsf{e}^{I} \mathsf{e}^{I \dagger} \rangle \mathsf{L} \right] = \mbox{Tr} \left[ \mathsf{E}^{I} \mathsf{L} \right]
\mbox{.}
\end{equation}

$\mathsf{E}^{I}$ is Hermitian, and contains the correlations between the Cartesian components of the incident field at the locations of the dipoles. (\ref{A24}) gives the expectation value of the absorbed power in terms of the state of coherence of the incident field and a response matrix $\mathsf{L}$.  Not only is (\ref{A24}) computationally convenient, it is conceptually important. The trace of the product of a matrix with the adjoint of another is an inner product in the abstract vector space of complex matrices. (\ref{A24}) therefore describes the projection of the state of coherence of the incident field onto the state of coherence of the field to which the system is responsive. In the case of continuous absorbers, an integral formulation of the same concept is available \cite{With_a}. $\mathsf{E}^{I}$ and $\mathsf{L}$ are Hermitian, and can be diagonalized: $\mathsf{E}^{I} = \sum_{m} \mathsf{v}_{m} \alpha_{m} \mathsf{v}_{m}^{\dagger}$, $\mathsf{L} = \sum_{n} \mathsf{u}_{n} \beta_{n} \mathsf{u}_{n}^{\dagger}$, which can be substituted in (\ref{A24}) to reveal
\begin{equation}
\label{A25}
 \langle W \rangle =  \sum_{mn} \alpha_{m} \beta_{n} \left| \mathsf{v}_{m}^{\dagger} \mathsf{u}_{n} \right|^{2}
\mbox{.}
\end{equation}
(\ref{A25}) describes the way in which the modes of the incident field $\mathsf{v}_{m}$ project onto the modes of the system, $\mathsf{u}_{n}$. The $\alpha_{m}$ are proportional to the levels of excitation of each of the incident modes, whereas the $\beta_{n}$ correspond to the relative responsivities with which the system absorbs energy in each of its modes. According to (\ref{A23}), the response matrix $\mathsf{L}$ contains two effects: the intrinsic ability of the dipoles to absorb energy, $ \mathsf{\Sigma}^{r}$, and the appearance of collective behaviour due to scattering, $\left[ \mathsf{M} \right]^{-1}$.

The electromagnetic Green's dyadic induces a wide variety of coherent behaviour, but additionally an elastic Green's dyadic could have been included in (\ref{A1}) to describe acoustic coupling through the excitation of surface waves.
The dynamical modes of interest might then be mediated by long-range phonon exchange. The overall principle, however, remains the same. Also, because DDA has proven to be an effective tool for modelling electromagnetic scattering from complex objects, including internal retardation, it use here for describing EAI means that the interferometric response of macroscopic inhomogenious objects, such as particulate material or droplets of a liquid, can be calculated. In this case, each macroscopic object is represented by a large collection of dipoles, rather than only having one dipole per object.

\subsection{Interferometry} \label{sec:interferometry}

In an experiment, the SUT is illuminated by a pair of phase-locked sources, and the absorbed power recorded as the differential phase between the sources is rotated. The absorbed power displays a fringe, and the complex visibility of this fringe is measured. In reality, the frequencies of the sources can be offset slightly, and the amplitude and phase of the fringe measured directly by using lock-in techniques. If the complex visibility is recorded for different pairs of source locations, either in the near field or far field, or a mixture of both, the response matrix $\mathsf{L}$ can be found.

Suppose that two sources are used, one producing the incident field $\mathsf{e}^{I}_{1} e^{j \phi_{1}}$, and the other producing $\mathsf{e}^{I}_{2} e^{j \phi_{2}}$. According to (\ref{A23}), when both sources illuminate the sample simultaneously,
\begin{equation}
\label{B1}
W =  \left[ \mathsf{e}^{I \dagger }_{1} e^{-j \phi_{1}} +  \mathsf{e}^{I \dagger }_{2} e^{-j \phi_{2}} \right] \mathsf{L} \left[ \mathsf{e}^{I}_{1}  e^{j \phi_{1}} +  \mathsf{e}^{I}_{2}  e^{j \phi_{2}} \right]
\mbox{.}
\end{equation}
Ensemble averaging is not required because the sources are fully coherent. (\ref{B1}) has four terms,
\begin{eqnarray}
\label{B2}
W & = &  \mathsf{e}^{I \dagger }_{1}  \mathsf{L}  \mathsf{e}^{I}_{1} +
\mathsf{e}^{I \dagger }_{2}  \mathsf{L}  \mathsf{e}^{I}_{2} +
  \mathsf{e}^{I \dagger }_{1}  \mathsf{L}  \mathsf{e}^{I}_{2} +
\mathsf{e}^{I \dagger }_{2}  \mathsf{L}  \mathsf{e}^{I}_{1} \\ \nonumber
 & = & W_{11} + W_{22} +  W_{12} e^{j \Delta \phi}+ W_{21} e^{-j \Delta \phi}
\mbox{,}
\end{eqnarray}
where $\Delta \phi = \phi_{2}-\phi_{1} $ is the phase difference between the sources. The first two terms, $W_{11}$ and $W_{22}$, are real, and give the powers dissipated by the sources individually. The second two terms are complex conjugates of each other $W_{12} = W_{21}^{\ast}$, and produce a fringe when the differential phase is rotated. It follows that $W_{11}, W_{22}, W_{21}, W_{12}$ can be found experimentally. If the field vectors  $\mathsf{e}^{I}_{1} e^{j \phi_{1}}$ and  $\mathsf{e}^{I}_{2} e^{j \phi_{2}}$ corresponded to point sources at the positions of the dipoles, the  response matrix $\mathsf{L}$ could be found directly, but in reality only distributed fields are possible.

Although the response matrix can be computed for different forms of illumination, plane waves, spherical waves, Gaussian beams, etc., here we assume that point sources, in the form of miniature current probes, are used. The current density
of a point source at ${\bf r}^{P}$ is
\begin{equation}
\label{B3}
{\bf J}^{P}({\bf r}') = \delta( {\bf r}' - {\bf r}^{P}) {\bf I}^{P}
\mbox{,}
\end{equation}
where ${\bf I}^{P}$ is the vector-valued current-moment. This source produces an incident field of the form
\begin{equation}
\label{B4}
{\bf E}^{I}({\bf r}) = \int \overline{\overline{\bf G}}^{P}( {\bf r}; {\bf r}' ) \cdot {\bf J}^{P}({\bf r}') \, d^{3}{\bf r}'
\mbox{,}
\end{equation}
where $\overline{\overline{\bf G}}^{P}( {\bf r}; {\bf r}' )$ is the Green's dyadic that couples the source current to the incident field at the dipoles. Substituting (\ref{B3}) in (\ref{B4}) leads to the discretized  form
\begin{equation}
\label{B5}
\mathsf{e}^{I} = \mathsf{G}^{P} \mathsf{i}^{P}
\mbox{.}
\end{equation}
For a single, simply-polarized source, $\mathsf{i}^{P}$ has one non-zero entry corresponding to the position and polarization of the probe. Using (\ref{B5}) and calculating the induced current moment, $\mathsf{i}^{I}$ at each dipole by integrating (\ref{A12}) gives
\begin{equation}
\label{B6}
\mathsf{i}^{I} =  \mathsf{\Sigma} \mathsf{M}^{-1}  \mathsf{G}^{P} \mathsf{i}^{P}
\mbox{.}
\end{equation}
The dissipated power can be written
\begin{eqnarray}
\label{B7}
 W  & =  & \frac{1}{2} \mbox{Re} \left[ \, \mathsf{e}^{\dagger} \mathsf{i}^{I} \, \right] \\ \nonumber
  & = & \mathsf{i}^{P \dagger} \mathsf{G}^{P \dagger} \left[ \mathsf{M}^{-1} \right]^{\dagger} \mathsf{\Sigma}^{r} \left[ \mathsf{M}^{-1} \right] \mathsf{G}^{P} \mathsf{i}^{P} \\ \nonumber
  & = &  \mathsf{i}^{P \dagger} \mathsf{G}^{P \dagger} \mathsf{L} \mathsf{G}^{P} \mathsf{i}^{P} \\ \nonumber
  & = &  \mathsf{i}^{P \dagger} \mathsf{H} \mathsf{i}^{P}
\mbox{.}
\end{eqnarray}
$\mathsf{L}$ is the intrinsic response matrix with respect to the incident field at the positions of the dipoles, whereas $\mathsf{H}$ is the response matrix with respect to current moments at the positions of the external sources.

To carry out interferometry, two sources are needed. Imagine that the current moment corresponding to source 1 is $\mathsf{i}^{P}_{1} e^{j \phi_{1}}$, and the current moment corresponding to source 2 is
$\mathsf{i}^{P}_{2} e^{j \phi_{2}}$. $\mathsf{i}^{P}_{1}$ and $\mathsf{i}^{P}_{2}$ are unit vectors
having a single real entry corresponding to a single current source with a single Cartesian component.
When both sources illuminate the sample simultaneously,
\begin{equation}
\label{B8}
W =  \left[ \mathsf{i}^{P \dagger }_{1} e^{-j \phi_{1}} +  \mathsf{i}^{P \dagger }_{2} e^{-j \phi_{2}} \right] \mathsf{H} \left[ \mathsf{i}^{P}_{1}  e^{j \phi_{1}} +  \mathsf{i}^{P}_{2}  e^{j \phi_{2}} \right]
\mbox{.}
\end{equation}
Ensemble averaging is not required because the sources are coherent. (\ref{B8}) has four terms
\begin{eqnarray}
\label{B9}
W & = &  \mathsf{i}^{P \dagger }_{1}  \mathsf{H}  \mathsf{i}^{P}_{1} +
\mathsf{i}^{P \dagger }_{2}  \mathsf{H}  \mathsf{i}^{P}_{2} +
  \mathsf{i}^{P \dagger }_{1}  \mathsf{H}  \mathsf{i}^{P}_{2} +
\mathsf{i}^{P \dagger }_{2}  \mathsf{H}  \mathsf{i}^{P}_{1} \\ \nonumber
 & = & H_{11} + H_{22} +  H_{12} e^{j \Delta \phi}+ H_{21} e^{-j \Delta \phi}
\mbox{.}
\end{eqnarray}
The first two terms, $H_{11}$ and $H_{22}$, are real, and give the powers dissipated by the sources individually. The second two terms satisfy $H_{12} = H_{21}^{\ast}$, and produce a fringe when the differential phase is rotated. Thus by measuring the power absorbed when each source is applied separately, and the amplitude and phase of the fringe when the two sources are applied simultaneously, $H_{11}, H_{22}, H_{21},$ and $H_{12}$ can be found.

Now move the two sources around within some sampling volume, stepping from one sample position to another. Usually the sample points will lie on a line or plane in the near field or far field, but more complicated paths can be used depending on the sampling needed. Construct the matrix $\mathsf{I}^{P}$, where column $n$, $\mathsf{i}^{P}_{n}$, corresponds to the source being at position $n$. Guided by (\ref{B7}), form the new matrix
\begin{equation}
\label{B10}
\mathsf{Q} = \mathsf{I}^{P \dagger} \mathsf{G}^{P \dagger} \left[ \mathsf{M}^{-1} \right]^{\dagger}
\mathsf{\Sigma} \left[ \mathsf{M}^{-1} \right] \mathsf{G}^{P} \mathsf{I}^{P}
\mbox{.}
\end{equation}
$\mathsf{Q}$ is not Hermitian, $\mathsf{\Sigma} \neq \mathsf{\Sigma}^{\dagger}$. Consider the Hermitian part of $\mathsf{Q}$,
\begin{eqnarray}
\label{B11}
\frac{1}{2}\left[ \mathsf{Q} + \mathsf{Q}^{\dagger} \right] & = & \mathsf{I}^{P \dagger} \mathsf{G}^{P \dagger} \left[ \mathsf{M}^{-1} \right]^{\dagger} \mathsf{\Sigma}^{r} \, \left[ \mathsf{M}^{-1} \right]  \mathsf{G}^{P} \mathsf{I}^{P} \\ \nonumber
 & = & \mathsf{I}^{P \dagger} \mathsf{G}^{P \dagger} \mathsf{L}  \mathsf{G}^{P} \mathsf{I}^{P} \\ \nonumber
 & = & \mathsf{I}^{P \dagger} \mathsf{H} \mathsf{I}^{P}
\mbox{.}
\end{eqnarray}
According to (\ref{B11}), the $n$'th element of the leading diagonal of $\mathsf{H}$ corresponds to the total power absorbed when single source $n$ illuminates the sample. The off-diagonal terms describe the fringe when a pair of phase coherent sources $n$ and $n'$ are used. It is therefore possible to populate $\mathsf{H}$ experimentally. If one source is held fixed at $n$, and the other moved through all other positions $n'$, the complex visibility of the spatial fringe is
\begin{equation}
\label{B12}
\gamma_{nn'} = \frac{2 H_{nn'}}{H_{nn} + H_{n'n'}}
\mbox{.}
\end{equation}
It is possible to plot out $| \gamma_{nn'} |$, with respect to reference point $n$, by moving along the $n$'th row of $\mathsf{H}$. These plots reveal the transverse and longitudinal coherence lengths, areas, and volumes of the polarized point-source excitation to which the structure is sensitive. $\mathsf{H}$ is the response matrix of the system with respect to point source excitation in the scanning volume. Once it has been populated experimentally, it can be diagonalized:
\begin{equation}
\label{B13}
\mathsf{H}   =   \mathsf{G}^{P \dagger} \mathsf{L}  \mathsf{G}^{P}
  =  \sum_{n} \mathsf{w}_{n} \gamma_{n} \mathsf{w}_{n}^{\dagger}
\mbox{.}
\end{equation}
Substituting (\ref{B13}) in the last line of (\ref{B7}), shows that if the source vector corresponds to each of the normalized eigenvectors in turn, $ \mathsf{i}^{P} = \mathsf{w}_{m}$, then the eigenvalues are the relative responsivities of the system. The $\mathsf{w}_{m}$ correspond to source vectors that excite modes of the system. Following (\ref{B6}), the dipole moment vector $\mathsf{p}_{m}$ of mode $m$ is
\begin{equation}
\label{B14}
\mathsf{p}_{m} = \left( \frac{1}{-i \omega} \right) \mathsf{\Sigma} \left[ \mathsf{M}^{-1} \right] \mathsf{G}^{P} \mathsf{w}_{m}
\mbox{.}
\end{equation}
No attempt is made to recover the displacements, as generally the charge at each site is not known.

Response matrix $\mathsf{H}$ combines three effects: (i) The generation of distributed fields by point current sources, described by $\mathsf{G}^{P}$; (ii) scattering interactions between the dipoles, described by $\left[ \mathsf{M}^{-1} \right]$; and (iii) the intrinsic ability of each dipole to absorb energy, described by $\mathsf{\Sigma}^{r}$. The second and third contributions, which generally cannot be separated, are intrinsic to the system, whereas the first term is a feature of the measurement. One could calculate
\begin{equation}
\label{B15}
\mathsf{e}_{m}^{P} =  \mathsf{G}^{P} \mathsf{w}_{m}
\mbox{,}
\end{equation}
and although these are field modes at the positions of the dipoles, they still relate to the original source volume. To calculate the modes at the positions of the dipoles, it is necessary to `deconvolve' the illumination patterns of the sources. Using (\ref{B13}), the deconvolved field modes are found by diagonalising
\begin{equation}
\label{B16}
\mathsf{L}  = \left[ \mathsf{G}^{P \dagger} \right]^{-1} \mathsf{H}  \left[ \mathsf{G}^{P} \right]^{-1}
=  \sum_{n} \mathsf{u}_{n} \beta_{n} \mathsf{u}_{n}^{\dagger}
\mbox{.}
\end{equation}
Generally, the Green's matrix will be singular, or poorly conditioned, depending on how well the sample points and polarizations span the vector space of fields at the dipoles. For poorly conditioned mappings, the generalized inverse should be used to recover those degrees of freedom that are available to the experiment. In reality, it is only necessary to include enough sample points to determine the significant degrees of freedom contained in $\mathsf{L}$, which may be fewer than the number of dipoles. One approach is to keep on adding, during an experiment, more and more points until the measurement has converged. Incremental Singular Value Decomposition can be used to diagonalise  $\mathsf{L}$ in real time until all of the available degrees of freedom have been found. Finally, the incident fields can be mapped onto the dynamical modes of the dipoles through
\begin{equation}
\label{B17}
\mathsf{p}_{m} = \left( \frac{1}{-i \omega} \right) \mathsf{\Sigma} \left[ \mathsf{M}^{-1} \right] \mathsf{z}_{m}
\mbox{.}
\end{equation}

\section{Simulations} \label{sec:simulations}
\begin{figure}[!]
\includegraphics[scale=0.8]{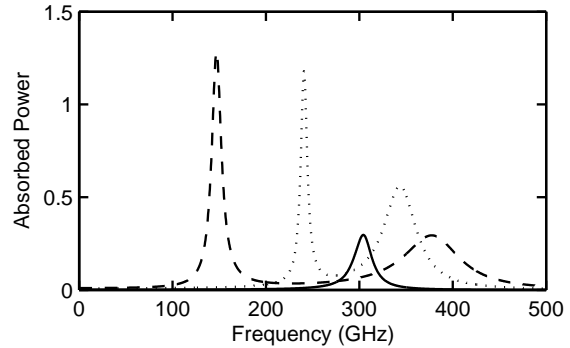}
\caption{\label{fig2} Total power absorbed (arbitrary units) as a function of frequency for two parallel dipoles, 0.1~mm apart, each having $\omega_{o}$ = 300~GHz, and $\Gamma$ = 20~GHz. The lines correspond to polarizabilities of (solid) $\alpha / \epsilon_{0}$ = 0.0005~mm$^{3}$, (dotted) $\alpha / \epsilon_{0}$ = 0.005~mm$^{3}$, (dashed) $\alpha / \epsilon_{0}$ = 0.01~mm$^{3}$.}
\end{figure}
A typical simulation of an experiment proceeds as follows. Establish a system of dipoles, and calculate the conductivity matrix $\mathsf{\Sigma}$ using (\ref{A11}). Calculate the scattering operator $\mathsf{M}$ using (\ref{A22b}), and invert it. Calculate $\mathsf{L}$ as defined by (\ref{A23}), and diagonalize it to find the eigenvalues and eigenvectors $\mathsf{u}_{n}$ and $\beta_{n}$. These are the collective dynamical modes of the system responsible for absorbing energy, which we would like to find. Choose a selection of source positions and polarizations, calculate the source Green's matrix $\mathsf{G}^{P}$, and the response matrix $\mathsf{H}$ through (\ref{B13}). Determine the amplitudes and phases that would be measured experimentally using (\ref{B7}), reverse the process using (\ref{B16}), and calculate the apparent dynamical modes using (\ref{B17}). Compare the actual modes with the recovered modes. Information may be lost due to the sampling used, the introduction of experimental errors, and the complete loss of certain fringe visibilities due to noise. During the work we found it instructive to produce `movies' showing the time evolution of the actual and recovered modes. It should be stressed that in all of the simulations that follow, only the total power absorbed by the complete system was regarded as measurable.

First, consider the resonant coupling of two parallel dipoles. Figure~\ref{fig2} shows the total power absorbed, in arbitrary units, as a function of frequency for two $z$-polarized dipoles, placed 0.1~mm apart on the $x$-axis, each having $\omega_{o}$ = 300~GHz, and $\Gamma$ = 20~GHz. The $z$-polarized source was located at $x$=$y$=10~mm to ensure that symmetric and antisymmetric modes could be excited. The lines correspond to different polarizabilities, thereby changing the degree of coupling, and the individual conductivities. The solid line shows the case when the polarizability is low, $\alpha / \epsilon_{0}$ = 0.0005~mm$^{3}$. The dipoles respond independently, giving rise to a single spectral feature at $300$~GHz, having a width of $\Gamma$ = 20~GHz. Because the conductivity is low, the absorbed power is small. The dotted line shows the case when the polarizability is increased to $\alpha / \epsilon_{0}$ = 0.005~mm$^{3}$. The dipoles interact, and split into an antisymmetric mode at 240~GHz, having a width of $\Gamma / 2$, and a symmetric mode at 340~GHz, having a width of $2 \Gamma $. The delocalized excitation is mediated mostly by near-field coupling, and the splitting is caused by a decrease (increase) in the effective spring constant of the antisymmetric (symmetric) mode as a consequence of the interaction. The absorbed power is larger than before because of the increased conductivity. The dashed line shows the case when the polarizability is high, $\alpha / \epsilon_{0}$ = 0.01~mm$^{3}$, and the splitting large. The relative strength of the peaks depends on the position of the source, indicating that spectral information alone can be misleading. Although not shown, when the polarizability is increased to $\alpha / \epsilon_{0}$ = 0.05~mm$^{3}$, the `high-conductivity' limit is approached, and the absorbed power falls to a very low level as the dipoles screen each other: the scattered field at each dipole is equal in magnitude and opposite in sign to the incident field.
\begin{figure}[!]
\includegraphics[scale=0.8]{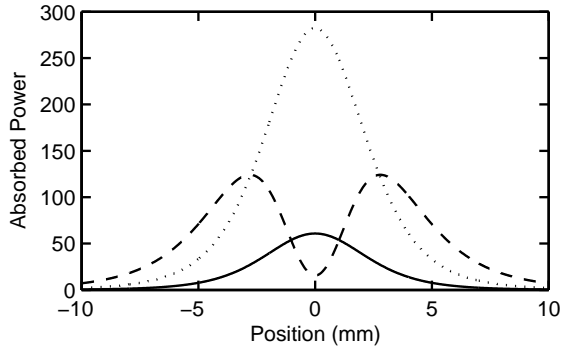}
\caption{\label{fig3} Total power absorbed (arbitrary units) as a single source is swept in the $x$-direction, at a distance $y=5$~mm from two parallel dipoles, 0.1~mm apart, each having $\omega_{o}$ = 300~GHz, $\Gamma$ = 20~GHz,
and $\alpha / \epsilon_{0}$ = 0.005~mm$^{3}$. The source frequencies were (dashed) 240~GHz, (solid) 300~GHz, and (dotted) 340~GHz.}
\end{figure}

Figure~\ref{fig3} shows the total power absorbed as a single source (using the near-field Green's dyadic) is swept in the $x$-direction, at a distance $y=5$~mm from the dipoles, with  $\alpha / \epsilon_{0}$ = 0.005~mm$^{3}$. According to Fig.~\ref{fig2}, the anti-symmetric mode peaks at 240~GHz, whereas the symmetric mode peaks at 340GHz. The plots in Fig.~\ref{fig3} correspond to source frequencies of (dashed) 240~GHz, (solid) 300~GHz, and (dotted) 340~GHz. Although the dipoles are not resolved spatially at this scan distance, the anti-symmetric mode has a null on-axis, whereas the symmetric mode has a peak on axis. From these absorption patterns one could infer the nature of the modes present at a single, intermediate frequency, say 300~GHz, which is the solid line in Fig.~\ref{fig3}, but what happens if wish to determine the modal content directly using only a single-frequency source. To this end, two sources were held at fixed positions (source 1 at $x=$ 0~mm, $y=$ 5~mm, and source 2 at $x=$ 5~mm, $y=$ 0~mm), and the differential phase rotated. The fringe visibility was recovered from the simulation, and used to assemble the $\mathsf{H}$ matrix, which was then converted to the $\mathsf{L}$ matrix by `deconvolving' the beam patterns of the sources, and then diagonalized to find the natural modes and their responsivities. The simulated experiment was repeated for the polarizabilities  and frequencies used in Figs.~\ref{fig2} and \ref{fig3}. In all cases, the spatial dynamical forms and responsivities of the anti-symmetric and symmetric modes recovered were precisely the same as those calculated directly. The same results were achieved regardless of whether the source probes were described by the near-field, far-field, or full Green's dyadics. At 240~GHz and $\alpha / \epsilon_{0}= $0.005~mm$^{3}$, the technique revealed a dominant anti-symmetric mode, corresponding to the peak in Fig.~\ref{fig2}. At 300~GHz, the modes were present in roughly equal amounts, whereas at 340~GHz, the symmetric mode was dominant. Crucially, the forms and relative responsivities of the two modes present at 300~GHz could be determined using a simulated interferometric measurement, even though the local absorption spectrum does not indicate the presence of two modes. In the case where $\alpha / \epsilon_{0}= $0.00005~mm$^{3}$, and over a wide range of frequencies, EAI correctly revealed the existence of two degenerate modes, corresponding to the oscillation of each dipole independently. Similar simulations were carried out for different source types and locations, and the dynamical forms and responsivities of the two modes could always be recovered. This simple example illustrates the nature of EAI, but care is required because the method is based on the ability to `deconvolve' the beam patterns of the sources, and so the accuracy of the final result will depend on the precision with which the experiment can be carried out.

\begin{figure}[!]
\includegraphics[scale=0.8]{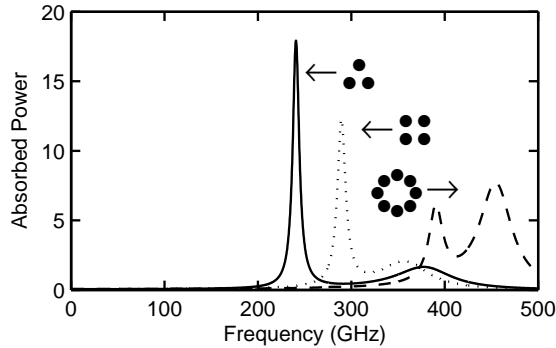}
\caption{\label{fig11} Total power absorbed (arbitrary units) as a function of frequency for a variety of ring-like configurations: (solid) 3 dipoles having $\alpha / \epsilon_{0}$ = 0.005~mm$^{3}$ placed 0.1~mm apart at the corners of a triangle; (dotted) 4 dipoles having $\alpha / \epsilon_{0}$ = 0.003~mm$^{3}$ placed 0.1~mm apart at the corners of a square; (dashed) 8 dipoles having $\alpha / \epsilon_{0}$ = 0.0004~mm$^{3}$ placed 0.0383~mm apart at the corners of an octagon. The power for the 8-dipole case was scaled by a factor of 10 before plotting.}
\end{figure}

Next, a series of simulations were performed on ring-like configurations of parallel dipoles. By using rings, boundary conditions at the ends are eliminated, and dispersion relationships can be found. Figure~\ref{fig11} shows the total absorbed power as a function of frequency for the following cases: (solid) 3 dipoles having $\alpha / \epsilon_{0}$ = 0.005~mm$^{3}$ placed 0.1~mm apart at the corners of a triangle; (dotted) 4 dipoles having $\alpha / \epsilon_{0}$ = 0.003~mm$^{3}$ placed 0.1~mm apart at the corners of a square; (dashed) 8 dipoles having $\alpha / \epsilon_{0}$ = 0.0004~mm$^{3}$ placed 0.0383~mm apart on the corners of an octagon. The polarizabilities were chosen to give similar spectral splittings, but the power for the 8-dipole case was scaled by a factor of 10 before plotting. The source was positioned at $x=$10~mm, $y=$0~mm, and operated in near-field mode.  The absorbed power revealed a low-level modulation as the source was rotated around the center at fixed distance, but the polar patterns are not shown for brevity. In all cases, the spectra acquired 2 peaks as a consequence of resonant interaction. Interactions take place not only around the ring, but also across the ring.

In the case of 3 dipoles, the peak at 378~GHz has a single dominant eigenvalue, corresponding to all dipoles oscillating in phase with equal amplitude. The peak at 240~GHz, is two-fold degenerate and accounts for the symmetric and antisymmetric motion of the 2 remaining degrees of freedom. In the case of 4 dipoles, the largest eigenvalue of the peak at 357~GHz corresponds to the collective delocalized in-phase oscillation of all dipoles, whereas the peak at 289~GHz is 2-fold degenerate and accounts for opposite pairs being out of phase. The system of 8 dipoles produces 2 peaks, with the one at 454~GHz being the in-phase collective oscillation of the whole system. The one at 390~GHz is two-fold degenerate, and corresponds to sinusoidal and cosinusoidal standing waves on the ring.

\begin{figure}[!]
\includegraphics[scale=0.8]{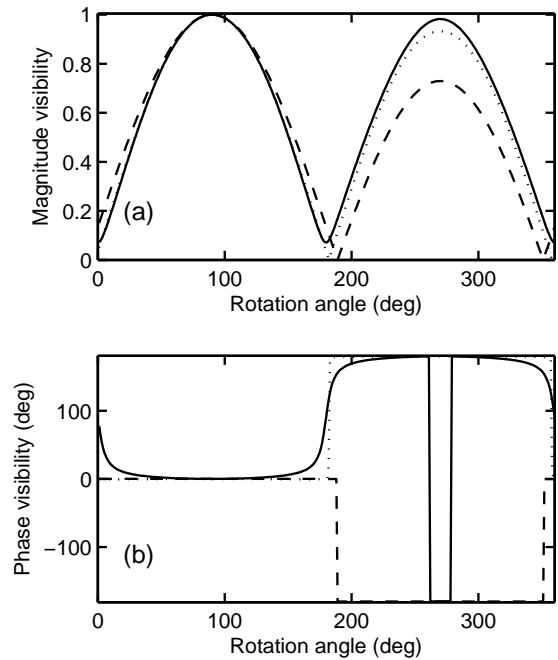}
\caption{\label{fig12} (a) Amplitude and (b) phase of the continuous fringe visibility of (solid) the triangular system at 240~GHz, (dotted) the square system at 289~GHz, and (dashed) the octagonal system at 390~GHz. The sources were swept around in a circle 5~mm from the centre. The reference angle was 90$^\circ$.}
\end{figure}

Even by comparing the full spectra in Fig.~\ref{fig11}, it is not possible to gain much information about the nature of the underlying modes. Moreover, the relative strengths of the peaks depends on the position of the source. If one is only interested in, and has access to, a limited frequency range, then there is little information in the polar power patterns that helps understand the number and forms of the modes. EAI can, however, reveal this information. To simulate EAI, two probes were rotated around the center of the system at a fixed distance of 5~mm, and the complex fringe visibility calculated for certain source pairs. Fig.~\ref{fig12} shows the continuous forms of typical visibilities. Remember that these are not spatial fringe patterns, but the complex visibility as a function of angle for a particular reference angle. The various lines are for the (solid) triangular, (dotted) square, and (dashed) octagonal arrangements describe previously. The reference angle was 90$^\circ$ in all cases, corresponding to a vertex point of the triangular and octagonal arrangements, and a face of the square arrangement. Figure~\ref{fig12} shows that the peak visibility is reduced when the sources are on opposite sides due to the `leading' edge screening the `trailing' edge. To recover the modes, a small number of sample points was used in each case; typically one more than the number of dipoles, avoiding the obvious degeneracies. It was found that the forms of the modes and their relative responsivities could be recovered easily at randomly chosen frequencies, without much care in selecting the source positions. Also, the recovery process depended little on whether the sources were described by the near-field, far-field, or full Green's dyadics. Of course, only the full Green's dyadic describes the physical reality, but the tolerance to the precise form of the illumination pattern indicates that, in principle, it should be possible to carry out interferometry using a variety of different near-field and far-field probes.

\begin{figure}[!]
\includegraphics[scale=0.8]{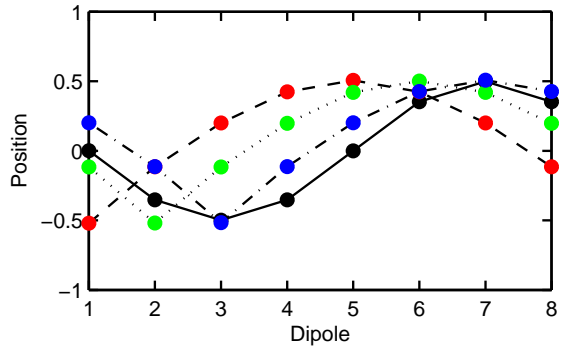}
\caption{\label{fig13} (Color online) Modal forms of the 413~GHz resonance of octagonal systems of dipoles with the resonant frequency of one dipole increased to 400~GHz. The reduced-mass dipole was moved around the chain (solid) none, (dashed) position 1, (dotted) position 2, (dot-dashed) position 3.}
\end{figure}

\begin{figure}[!]
\includegraphics[scale=0.8]{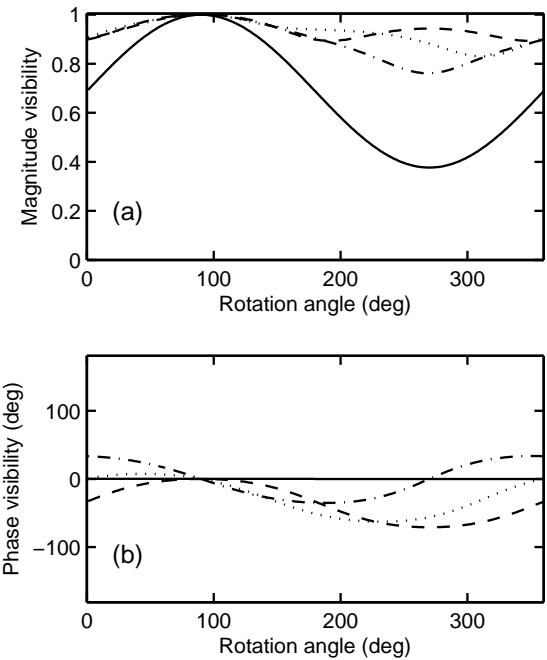}
\caption{\label{fig14} (a) Amplitude and (b) phase of the recorded fringe visibility at 413~GHz of octagonal systems, with the resonant frequency of one dipole changed to 400~GHz. The lines correspond to (solid) none, (dashed) position 1 (corresponding to the 90$^\circ$ reference angle), (dotted) position 2, (dot-dashed) position 3.}
\end{figure}

The octagonal system was considered again, but with the resonant frequency of one dipole increased to 400~GHz, corresponding to a reduction in the mass of the oscillating charge. Although not shown here, the principle effect on the spectrum is to split the original low-frequency line at 390~GHz into two separate frequencies, 387~GHz and 413~GHz, reflecting the breaking of the degeneracy of the sinusoidal and cosinusoidal modes. There is also a quasi-static mode at 290~GHz. The dominant modal forms are shown in Fig.~\ref{fig13}, at 413~GHz, as the reduced mass is moved around the octagon. In each case, the partner mode at 387~GHz has a standing wave that leaves the dipole with the reduced mass stationary. The corresponding visibility functions are shown in Fig.~\ref{fig14}, where the reference angle was 90$^\circ$. The visibility plots, particularly the phase, clearly show the movement of the defect dipole.

A number of discrete source positions where then used in a attempt to detect the movement of the defect dipole around the octagonal structure. The sample points were placed on a circle, 5~mm from the center, 20$^\circ$ apart, with the first one opposite a vertex. The modes and their relative responsivities could be recovered precisely regardless of whether the near-field or far-field source Green's dyadic was used. The forms of most high-order modes could be recovered precisely with 8 or greater sample points, with some low-level of aliasing with 5 to 7 sample points, and not at all with 4 sample points. The determination of the position of the defect dipole, depended little on the positions of the sample points. Overall, the modes could be recovered in detail despite the partial screening of the `dark side' of the ring.

\begin{figure}[!]
\includegraphics[scale=0.8]{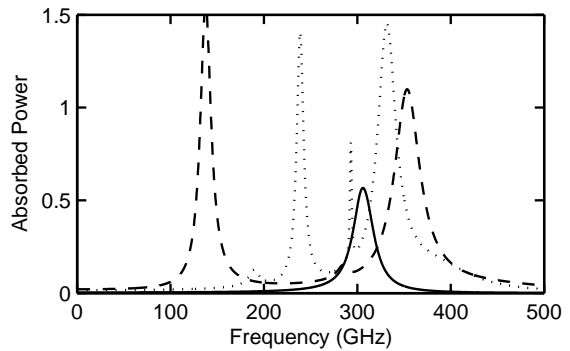}
\caption{\label{fig4} Total power absorbed (arbitrary units) as a function of frequency for a linear chain of 5 parallel dipoles, 0.1~mm apart, each having $\omega_{o}$ = 300~GHz, and $\Gamma$ = 20~GHz. The different lines correspond to polarizabilities, (solid) $\alpha / \epsilon_{0}$ = 0.0005~mm$^{3}$, (dotted) $\alpha / \epsilon_{0}$ = 0.005~mm$^{3}$, (dashed) $\alpha / \epsilon_{0}$ = 0.01~mm$^{3}$.}
\end{figure}

\begin{figure}[!]
\includegraphics[scale=0.8]{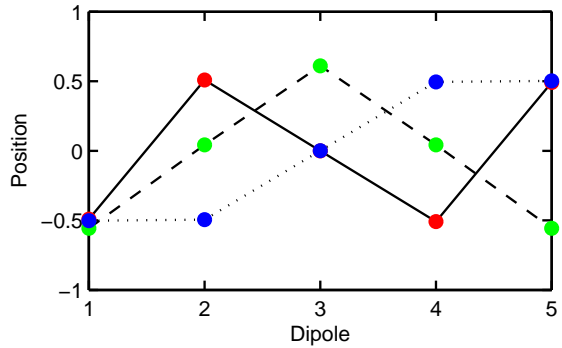}
\caption{\label{fig5} (Color online) Modal forms of the resonances at (solid) 239~GHz, (dashed) 293~GHz, and (dotted) 332~GHz in Fig.~\ref{fig4}. All dipoles are either perfectly in phase or out of phase. The lines are drawn to guide the eye only.}
\end{figure}

Turn to linear arrays. Figure~\ref{fig4} shows the total power absorbed, in arbitrary units, as a function of frequency for 5, $z$-polarized dipoles, placed 0.1~mm apart on the $x$ axis, each having $\omega_{o}$ = 300~GHz, and $\Gamma$ = 20~GHz. The $z$-polarized source was located at $x$=10~mm, $y$=10~mm to ensure that symmetric and antisymmetric modes could be excited. The solid line shows the case when the polarizability is low, $\alpha / \epsilon_{0}$ = 0.0005~mm$^{3}$, and the dipoles are largely, but not completely, decoupled. The dotted line shows the case when the polarizability is increased to $\alpha / \epsilon_{0}$ = 0.005~mm$^{3}$. The dipoles interact, and give rise to 5 spectral features at 187, 239, 293, 332, and 381~GHz; these alternate between symmetric and antisymmetric forms. It is not necessarily the case that the number of features is equal to or smaller than the number of dipoles, because for large systems spatially harmonic response due to widely separated dipoles may exist. The dashed line shows the case when the polarizability is increased further, $\alpha / \epsilon_{0}$ = 0.01~mm$^{3}$. The response becomes very similar to that of Fig.~\ref{fig2}, but now both modes are antisymmetric whereas before one was symmetric and the other antisymmetric. Figure~\ref{fig5} shows the forms of the dominating modes, for  $\alpha / \epsilon_{0}$ = 0.005~mm$^{3}$, at the resonance frequencies of (solid) 239~GHz, (dashed) 293~GHz, and (dotted) 332~GHz. Each filled circle shows the position of the displaced charge at $t=0$: $z_{i} (0)$. Dipoles having the same sign in the figure remain perfectly in phase as time progresses, whereas those having opposite sign remain perfectly out of phase. Those at zero remain at zero. The lines are drawn to guide the eye only. Table~\ref{tabb} gives the elements of the $\mathsf{H}$ matrix recovered from the fringes when the 5-dipole structure, centred on $x=$0.0~mm, was illuminated by 2 $z$-polarized sources at 280GHz, which according to the dotted line of Fig.~\ref{fig4} means that the sample was not resonant in any of its natural modes. A total of 5 source positions were used, $x=$1, 2, 3, 4, 5~mm, with $y=$ 5~mm in all cases. This set of data was then corrected for the beam patterns, and the dynamical modes calculated. The eigenvalues of the recovered modes are given in the next to last line of Table~\ref{tabb}. The forms of the recovered modes were then reconstructed and compared with the modes associated with the peaks in Fig.~\ref{fig4}, 3 of which are shown in Fig.~\ref{fig5}. The corresponding identifications are listed on the last line of Table~\ref{tabb}, which are reasonable given the source frequency, and the frequencies of the resonances. The positions of the test sources were moved randomly, and it was always found that 5 source positions were needed to recover all of the dynamical modes accurately. If more than 5 were used, the recovered modes and eigenvalues did not change, as would be expected for a well-conditioned problem. The conclusion is that a simple interferometric measurement, at a single frequency, based on measuring the total power absorbed, has enabled the forms and responsivities of the dynamical modes to be found.

\begin{table}
\begin{center}
 \begin{ruledtabular}
\begin{tabular}{c c c c c}
109.07 &  81.90 &  53.10 &  31.51 &  17.87 \\
81.90 &  64.16 &  43.80 &  27.73 &  17.05 \\
53.10 &  43.80 &  31.68 &  21.39 &  14.11 \\
31.51 &  27.73 &  21.39 &  15.40 &  10.80 \\
17.87 &  17.05 &  14.11 &  10.80 &   7.99 \\ \hline
0.00 &  95.05 & -116.54 &  74.37 & -62.26 \\
-95.05  &  0.00 & 148.46 & -20.51 & -156.98 \\
116.54 & -148.46 & 0.00 & -168.93 &  54.64 \\
-74.37 &  20.51 & 168.93 &  0.00 & -136.44 \\
62.26 & 156.98 & -54.64 & 136.44 &  0.00 \\ \hline
0.331 & 0.048 & 0.014 & 0.013 & 0.003 \\
293 & 239 & 332 & 187 & 381 \\
\end{tabular}
\end{ruledtabular}
\caption{\label{tabb} Fringe parameters of the 5 dipole structure described in the text.  Two $z$-polarized, 280~GHz,
sources were placed in pairs at the positions $x=$1.0, 2.0, 3.0, 4.0, and 5.0~mm, with $y=$ 5.0~mm in all cases. The top block of data gives the amplitudes of the elements of the $\mathsf{H}$ matrix, and the middle block gives the corresponding phases. All values were recovered from the `measured' fringes. The penultimate line gives the eigenvalues of the natural modes at the dipoles, and the bottom line gives the frequency (GHz) at which the corresponding mode is resonant.}
\end{center}
\end{table}

\begin{figure}[!]
\includegraphics[scale=0.8]{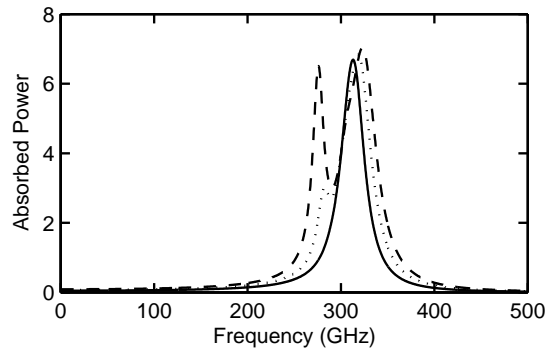}
\caption{\label{fig6} Total power absorbed (arbitrary units) as a function of frequency for 11 parallel dipoles,
0.1~mm apart, each having $\omega_{o}$ = 300~GHz, and $\Gamma$ = 20~GHz. All of the dipoles except the central one had a polarizability of $\alpha / \epsilon_{0}$ = 0.0005~mm$^{3}$, whereas the central one had (solid) $\alpha / \epsilon_{0}$ = 0.001~mm$^{3}$, (dotted) $\alpha / \epsilon_{0}$ = 0.003~mm$^{3}$, and (dashed) $\alpha / \epsilon_{0}$ = 0.005~mm$^{3}$.}
\end{figure}

\begin{figure}[!]
\includegraphics[scale=0.8]{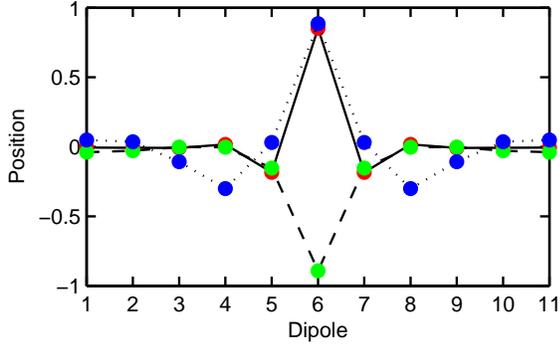}
\caption{\label{fig7} (Color online) Modal forms of the strongest absorptive modes at (solid) 276~GHz, (dashed) 323~GHz, and (dotted) 300~GHz of the dashed spectrum shown in Fig.~\ref{fig6}. All dipoles are either perfectly in phase or perfectly out of phase, and the lines are drawn to guide the eye only.}
\end{figure}

Figure~\ref{fig6} shows the total power absorbed, in arbitrary units, as a function of frequency for 11, $z$-polarized dipoles, placed 0.1~mm apart, each having $\omega_{o}$ = 300~GHz, and $\Gamma$ = 20~GHz. All of the dipoles except the central one had a polarizability of $\alpha / \epsilon_{0}$ = 0.0005~mm$^{3}$, whereas the central one had (solid) $\alpha / \epsilon_{0}$ = 0.001~mm$^{3}$, (dotted) $\alpha / \epsilon_{0}$ = 0.003~mm$^{3}$, and (dashed) $\alpha / \epsilon_{0}$ = 0.005~mm$^{3}$. The $z$-polarized source was located at $x$=0 and $y$=1~mm, and operated in the near-field mode. Even when the polarizability is low, there is evidence of interaction, because the peak is at 310~GHz, not 300~GHz. As the polarizability of the central dipole is increased, the spectrum splits into strong features at 276~GHz and 323~GHz, indicating the excitation of two resonant modes. Figure~\ref{fig7} shows the form of the strongest mode at each of 276, 300, and 323~GHz. At 276~GHz, the highly polarized dipole induces a localized excitation, with the nearest neighbors 180$^{\circ}$ out of phase with the central dipole. At 323~GHz, the mode is still localized, but now the nearest neighbors are in phase with the central, exciting dipole. There is no spectral feature at 300~GHz, and the dominant mode is less localized, with the nearest neighbors being stationary, and the next nearest neighbors being out of phase. The excitation is influenced by the boundary conditions at the ends of the chain. The modal forms change significantly over the 280-320~GHz range, and can take on forms where the central dipole launches traveling, albeit rapidly decaying, waves. Beating can also be seen between different spatial frequencies.

\begin{figure}[!]
\includegraphics[scale=0.8]{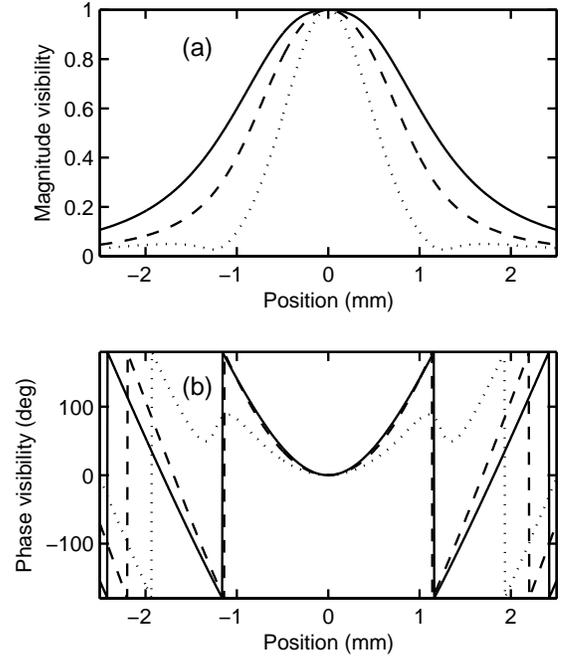}
\caption{\label{fig8} (a) Amplitude and (b) phase of the recorded fringe visibility at frequencies of (solid) 276~GHz, (dashed) 323~GHz, and (dotted) 300~GHz. The sources were swept along a line, 1~mm from the chain of dipoles, and the visibilities are referenced to the central position.}
\end{figure}

Figure~\ref{fig8} shows the amplitudes and phases of the continuous visibilities when two sources are swept along a parallel line 1~mm from the chain. The sources were operated in near-field mode, and the plots correspond  to frequencies of (solid) 276~GHz, (dashed) 323~GHz, and (dotted) 300~GHz. The fringes give a clear indication of the changing form of the localized mode. A set of visibilities were recorded with only 11 source positions, at the same distance and spanning the spatial range shown in Fig.~\ref{fig8}. It was found that at each frequency the forms of all of the dynamical modes and their responsivities could be recovered, regardless of whether the probes were operated in the near-field or far-field mode. As the total number of source positions was increased from 2, the recovered modes were in error until 11 source positions were used, at which point all of the modal forms were found precisely. As the number of source positions increased above 11, no change was seen, showing that convergence could be reached, and recognized.

\begin{figure}[!]
\includegraphics[scale=0.8]{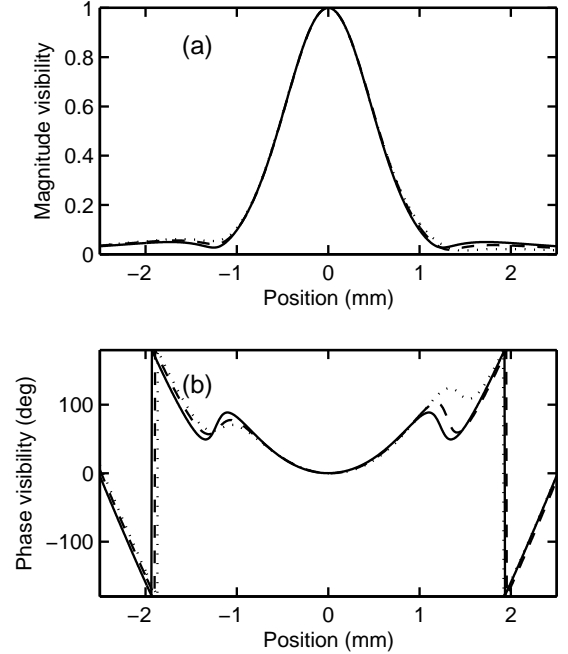}
\caption{\label{fig9} (a) Amplitude and (b) phase of the recorded fringe visibility at 300~GHz. A highly polarizable dipole was placed at (solid) $x=$0.0~mm, (dashed) $x=$0.1~mm, (dashed) $x=$0.2~mm. The sources were swept along a line, 1~mm from the chain of dipoles, and the visibilities are referenced to the central position.}
\end{figure}

Figure~\ref{fig9} shows fringe visibilities at 300~GHz, when the position of the highly polarizable dipole was shifted one place at a time: (solid) $x=$0.0~mm, (dashed) $x=$0.1~mm, (dashed) $x=$0.2~mm. The sources were swept along a line 1~mm from the chain of dipoles, and the visibilities referenced to the central position. The fringe is sensitive to the position of the localized mode. Moreover, all of the modal forms could again be recovered with 11 source positions, allowing the position of the highly polarisable dipole to be located.

\begin{figure}[!]
\includegraphics[scale=0.8]{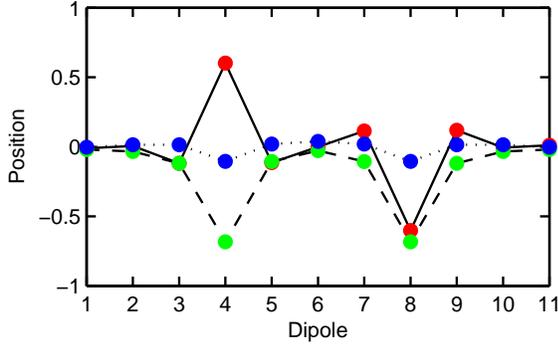}
\caption{\label{fig10} (Color online) Modal forms of the strongest absorptive modes at (solid) 270~GHz, (dashed) 330~GHz, and (dotted) 284~GHz of a system comprising 11 dipoles. All of the dipoles had a polarizability of $\alpha / \epsilon_{0}$ = 0.0005~mm$^{3}$, except the ones at $x=$-0.2~mm and $x=$+0.2~mm (dipoles 4 and 8), which had a polarizability of $\alpha / \epsilon_{0}$ = 0.005~mm$^{3}$.}
\end{figure}

More complicated linear systems were then considered. For example, Fig.~\ref{fig10}  shows the modal forms of two interacting localized excitations. All of the dipoles had a polarizability of $\alpha / \epsilon_{0}$ = 0.0005~mm$^{3}$, except the ones at $x=$-0.2~mm and $x=$+0.2~mm (dipoles 4 and 8), which had a polarizability of $\alpha / \epsilon_{0}$ = 0.005~mm$^{3}$. The two highly polarized dipoles now oscillate either in phase or out of phase, with the interaction mediated by the background chain. Again, we were able to recover all of the modes and responsivities at any frequency using EAI.

\begin{figure}[!]
\includegraphics[scale=0.8]{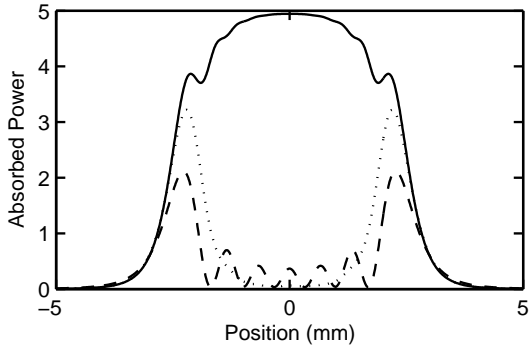}
\caption{\label{fig16} Total power absorbed (arbitrary units) as a single probe is scanned along a line parallel to a linear chain of 21 dipoles placed 0.2~mm apart, having $\omega_{0}=$300~GHz, $\Gamma =$20~GHz, and $\alpha / \epsilon_{0}=$0.5~mm$^3$ The scan distance was 1~mm. The lines correspond to frequencies of (dotted) 42~GHz, (dashed) 205~GHz, and  (solid) 500~GHz: the powers have been divided by 5$\times$10$^8$, 1$\times$10$^4$, and 1$\times$10$^7$ respectively before plotting.}
\end{figure}

\begin{figure}[!]
\includegraphics[scale=0.8]{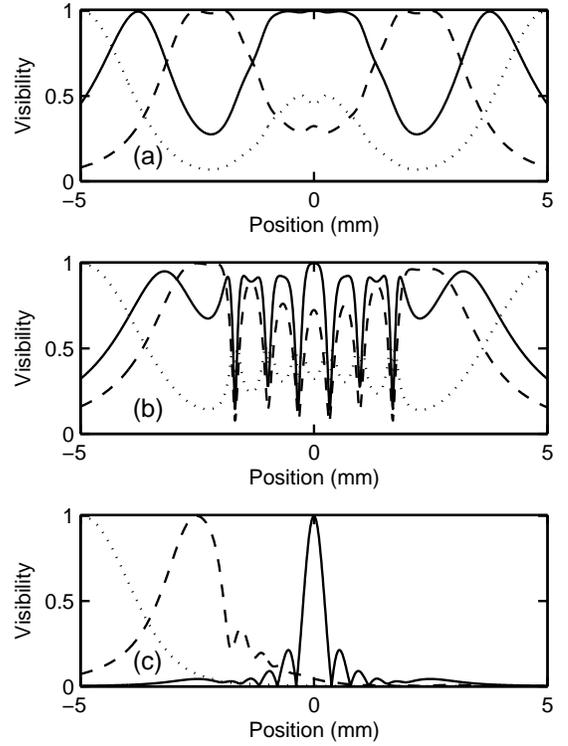}
\caption{\label{fig17} Amplitude of the continuous fringe visibility for an array of 21 dipoles placed 0.2~mm apart, $\omega_{0}=$300GHz, and $\Gamma =$20~GHz. (a) 42~GHz, (b) 205~GHz, (c) 510~GHz. The reference positions are (solid) 0~mm, (dashed) -2.5, and (dotted) -5.0~mm in all cases. The ends of the chain are at $\pm$2~mm.}
\end{figure}

Finally consider a linear array of 21 dipoles, placed 0.2~mm apart, having $\omega_{0}=$300GHz, $\Gamma =$20~GHz,
and $\alpha / \epsilon_{0}=$0.5~mm$^3$. The spectrum, not shown, has high-Q resonances at 42~GHz and 205~GHz, each of which has a single dominant eigenvalue. Figure~\ref{fig16} shows the power absorbed as a single probe was swept parallel to the chain, 1~mm from it, thereby producing a near-field scan. The two sweeps at (dotted) 42~GHz and (dashed) 205~GHz, show strong evidence of end effects, whereas the sweep at (solid) 500~GHz, where no spectral features are present, has a form typical of a continuous nanowire or sheet of resistive material. Figure~\ref{fig17} shows the amplitude of the fringe visibility at (top) 42~GHz, (middle) 205~GHz, (bottom) 500~GHz. The reference positions are (solid) 0~mm, (dashed) -2.5, and (dotted) -5.0~mm in all cases. The ends of the chain are at $\pm$2~mm.

\begin{figure}[!]
\includegraphics[scale=0.8]{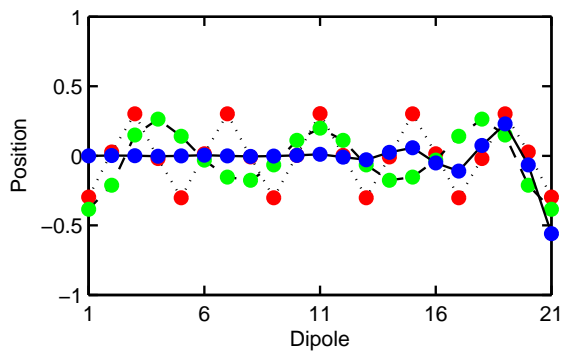}
\caption{\label{fig18} (Color online) Modal forms of the strongest absorptive modes at (dotted) 42~GHz, (dashed) 205~GHz, and (solid) 500~GHz of a linear chain of 21 dipoles having placed 0.2~mm apart, having $\omega_{0}=$300GHz, $\Gamma =$20~GHz, and $\alpha /\epsilon_{0}=$0.5~mm$^3$. The lines are drawn to guide the eye only.}
\end{figure}

Using pairs of dipoles at the same scan distance, we recovered the dynamical modes using EAI, and compared them with those calculated directly. A limited number of test-source positions, typically 20, were used, but no differences were seen when over 1000 were used. The behaviour at 42~GHz could always be recovered easily, and the behaviour at 205~GHz and 510~GHz could be recovered for linear scans $<$1~mm from the chain. At greater distances, recovery became uncertain, because of difficulty exciting dipoles near the ends. By always placing 2 sources in line with the dipoles, but beyond the ends, the problem could be solved. The situation is similar to edge absorption in continuous films. Figure~\ref{fig18} shows the recovered modes  in those cases where sufficient source positions were used to recover the modes accurately. The mode at (dotted)  42~GHz is a quasi-static solution, dominated by near-field coupling; where half of the dipoles are stationary, and the other half oscillate symmetrically, in perfect antiphase, about the centre of the chain. The fringe visibility, plotted in the top pane of Fig.~\ref{fig17}, shows directly that dipole interaction is strong enough to produce highly coherent extended behaviour along the whole length of the structure. Near-field interactions on linear chains are being considered for possible application as sub-wavelength waveguides. The mode at (dashed) 205~GHz is mediated by far-field retarded interaction. The dipoles at the ends are lightly bound, because they only have neighbors on one side, and oscillate with high amplitude; they form waves that travel inwards and interfere to produce a standing wave at the centre. The visibility plots shown in the middle pane of Fig.~\ref{fig17} reveal the existence of this large-scale collective behaviour. The bottom pane of Fig.~\ref{fig17} shows the visibility at 500~GHz, where no obvious spectral features are present. The two largest eigenmodes are degenerate, each corresponding to a high amplitude end wave that travels towards the centre: only one of them is shown in Fig.~\ref{fig18} (solid). Crucially, however, the effects at opposite ends are not coupled, as they are at 205~GHz. This behaviour is apparent in the visibility plots. Near the centre, the system behaves as an almost incoherent absorber, whereas at the ends, highly coherent absorption is seen. The absorption process at one end is incoherent with respect to the absorption process at the other. Compare the bottom pane of Fig.~\ref{fig17} with the middle and top panes. The various forms seen are typical of thin-film planar absorbers, and therefore the linear chain is becoming large enough to take on the characteristics of a continuous absorber, and macroscopic multipolar expansions can be used to describe behaviour \cite{Thomas_a}.

Many other simulations were performed during the course of the work. For example, we looked at arrangements where a single, lightly coupled dipole was placed next to a ring-like or linear chain, and found that we were able to separate out the behaviour of the lightly bound dipole from the collective behaviour of the chain. Also, we could identify the existence of `grain-boundary-like' defects on long chains.

\section{Practical Considerations} \label{sec:practical}

Although this paper focuses on the theory of EAI, it beneficial to comment on its practical application. The central requirements are (i) the ability to phase-lock two sources, (ii) the ability to scan the sources within a near-field or far-field volume, (iii) the ability to measure the total power absorbed, or at least the part of it that constitutes the operation of the device. Because the key elements are so simple, the scheme can be used at any wavelength. Crucially, high-power sources can be used, perhaps at low temperatures, extending the method well beyond anything that can be achieved by measuring the correlations in thermally radiated fields \cite{Agarwal_a, Carmin_a, Shch_a}.

It is certainly straightforward to assemble phased-locked sources for radio, infrared, and optical wavelengths. In fact, we have already demonstrated that it is possible to measure the state of coherence of the reception patterns of multimode power detectors over the range 195-270~GHz using absorbed power only \cite{Thomas_b}. In that work, the frequencies of the two sources were offset slightly to produce a time varying output that contained the fringe information in real time. In some cases, a single-frequency source would be used, say in the case of a laser and power splitter, and then $\lambda /4$ path-length switching would be needed to determine the quadrature components of the fringe.

There are many ways in which the SUT could be illuminated with different basis sets to span the dynamical modes of interest. For example, far-field probes in the form of optical fibres, or near-field probes in the form of coaxial AFM-like tips, could be scanned across or rotated around the SUT. In magnetic measurements, coils could be swept around and scanned across the surface of interest. In general, polarisation switching is needed, but this can be done in a straightforward manner. A crucial question is `how many samples must be taken to gain access to the degrees of freedom of interest?' We have already identified efficient real-time techniques for adding in more and more sample points until convergence is reached, and the significant degrees of freedom identified. EAI is, in general terms, quite similar to Aperture Synthesis Interferometry, used extensively in astronomy, and many of the data processing techniques developed for astronomical imaging, such as closure phase, could be incorporated here.

The final requirement is a way of determining the total power absorbed. In many applications, such as photovoltaics and optical sensors, the output of the device is already a direct measure of the power being absorbed through the degrees of freedom of interest.  In these cases, one might not even be interested in the surface physics per se, but a detailed characterisation of the nature, number, and responsivities of the electromagnetic modes to which the device is sensitive. Additionally, representative models based on DDA can be used to understand how the electromagnetic modes relate to the dynamical behaviour of the absorbing surface. In some cases, such as patterned films or biomolecules on surfaces, a way of monitoring the total power absorbed is required. Our own preferred approach is to attach the SUT to a micromachined SiN membrane, which also supports a thin-film temperature sensor. Such devices have been developed extensively for astronomical detectors, and when used at low temperatures can achieve extreme sensitivity \cite{Goldie_a, Goldie_b}. Our detectors \cite{Goldie_c} can detect absorbed powers of much less than 1~fW with noise levels of 1x10$^{-19}$~WHz$^{-1/2}$, making extreme thermometry of thin-film structures entirely possible.

\section{Conclusions}

We have shown that by illuminating an energy-absorbing surface or structure with a pair of phase-locked sources, it is possible to measure the state of coherence of the electromagnetic field to which the structure is sensitive. The tensor that describes the coherence can be diagonalized to find the amplitude, phase, and polarization patterns of the individual electromagnetic modes through which the structure can absorb power. In some cases, such as photovoltaics and detectors, this information is already sufficient to understand and characterize behaviour. It is also possible, however, to map the electromagnetic modes back onto the surface, or object, to gain representations of the dynamical modes responsible for absorption. These modes are intimately related to the existence of coherent phenomena, such as elastic waves and surface plasmons. In a continuous formulation of the method, not yet published, we have shown that it is possible to detect the presence of subsurface voids. The method can be implemented easily, and because high-power sources are used, can far outperform any experiment that relies on measuring the correlations in themally radiated fields. More speculatively, the use of high power sources, together with a space-frequency formulation, may open the way to studying non-linear surface effects.

\begin{acknowledgments}

This work was carried out while Prof. Withington was a Visiting Fellow at All Souls College, Oxford. Prof. Withington would like to thank the Warden and Fellows for their generosity and hospitality during his stay.

\end{acknowledgments}

\end{document}